\documentclass[pra,aps,10pt,twocolumn,showpacs,amsmath,amssymb]{revtex4-1}

\usepackage[english]{babel}
\usepackage{float}
\usepackage{afterpage}
\usepackage{graphicx}
\usepackage{dcolumn}
\usepackage{bm}
\usepackage{amsmath}
\usepackage{epsf}
\usepackage{epsfig}
\usepackage{amssymb}
\usepackage{color}

\usepackage{slashed}
\usepackage{latexsym}

\begin{document}
\title{Symmetries in nonlinear Bethe-Heitler process}
\author{K. Krajewska}
\email[E-mail address:\;]{Katarzyna.Krajewska@fuw.edu.pl}
\author{J. Z. Kami\'nski}
\affiliation{Institute of Theoretical Physics, Faculty of Physics, University of Warsaw, Ho\.{z}a 69,
00-681 Warsaw, Poland}
\date{\today}

\begin{abstract}
Nonlinear Bethe-Heitler process in a bichromatic laser field is investigated using strong-field 
QED formalism. Symmetry properties of angular distributions of created $e^-e^+$ pairs are analyzed. 
These properties are showed to be governed by a behavior of the vector potential characterizing the laser field,
rather than by the respective electric field component.
\end{abstract}


\maketitle

Quantum electrodynamics (QED) predicts a possibility of spontaneous emission of electron-positron pairs 
from a vacuum in the presence of a static electric field~\cite{Sauter,Schwinger}; the mechanism known as the 
Sauter-Schwinger mechanism (see, e.g., Ref.~\cite{Ehlotzky}). 
This nonlinear effect was also considered for time-dependent electric fields~\cite{Brezin,Popov}.
In the latter case, the pairs are tunneled through the QED vacuum by a strong low-frequency 
oscillating field; the point is that when an alternating electric field changes slowly over the Compton length
then it can be treated quasistationary. 
In the quasistationary picture, the pairs are produced if work performed by the electric field
over the Compton length provides the energy of $2m_{\rm e}c^2$, with the probability determined by the maximum 
electric field ${\cal E}$ (above, $m_{\rm e}$ is the electron mass). More specifically, the probability of pair 
creation scales as $\exp(-\pi {\cal E}_{\rm cr}/{\cal E})$ for ${\cal E}\ll {\cal E}_{\rm cr}$, which is the Sauter-Schwinger formula with 
the critical electric field ${\cal E}_{\rm cr}=1.3\times 10^{16}$ V/cm.

In this paper, another scenario of pair creation, the so-called nonlinear Bethe-Heitler process, in 
which electron-positron pairs are created in collisions of a super-intense laser beam with a beam 
of relativistic targets is investigated. We will show that in the case under investigations,
it is the vector potential (not the electric field) which determines properties of the  
process. This is particularly interesting in the context of Sauter-Schwinger mechanism.
In our case, even if we deal with a low-frequency laser field, in the reference 
frame of a target particle the field can exhibit, in fact, very rapid oscillations over the Compton length.
Let us remind that in usual experiments, incident targets move with very high kinetic energies; in
the SLAC experiment reaching almost 50 GeV that corresponds to the Lorentz factor $\gamma\simeq 10^5$~\cite{Bamber,Bula}.
Therefore for a typical Ti:Sapphire laser field, the Doppler upshifted frequency of the laser field in 
the target-particle rest frame of reference can reach even $m_{\rm e}c^2$. Also the laser field
strength in the moving reference frame will be significantly enhanced. In particular,
for record laser fields with intensities of the order of $(10^{20}-10^{22})$ W/cm$^2$~\cite{Mourou}, the 
corresponding electric fields measured in the moving frame of reference can even exceed ${\cal E}_{\rm cr}$.

In the context of the Sauter-Schwinger mechanism, we investigate the role of an incident laser
field in the nonlinear Bethe-Heitler process. This will be done for parameters available today in experiments, 
that may lead to considerable number of $e^-e^+$ pairs. Contrary however to experiments that have used electrons
as target particles (see, for instance, in Refs.~\cite{Bamber,Bula}), we will use protons. The point being that in this case one does not have to
account for the exchange diagram when calculating the $S$-matrix amplitude, which makes it more straightforward to interpret. For our purpose, the angular
spectra of product particles will be analyzed. The key question is how these distributions 
correlate with symmetry of both the electric field and the vector potential characterizing 
the incident laser field.

We consider the electron-positron pair creation in collisions of a proton with a bichromatic laser field 
using the theory and numerical methods developed in our most recent paper~\cite{b1}. Let us mention
that the theory presented there, and the current results as well, account for the proton recoil. 
We consider the case when both components of the laser field have commensurate 
frequencies and they are linearly polarized, with the polarization vectors along the $x$ axis. 
We also assume that the laser field propagates in the opposite $z$ direction. Even though the theory introduced in Ref.~\cite{b1}
is very general, we present the results which relate to the reference frame where the 
proton is at rest long before the collision takes place. In other words, if 
$p_{{\rm e}^-}$ and $p_{{\rm e}^+}$ stand for the four-momenta of created electron
and positron, respectively, whereas $q_{\rm i}$ and $q_{\rm f}$ are the incoming and outgoing
four-momenta of a proton, all numerical results will relate to the reference frame
in which ${\bm q}_{\rm f}={\bm 0}$. Consequently, $\omega$ will represent a Doppler-upshifted 
fundamental frequency of the laser field oscillations. The same approach was used in Ref.~\cite{b1}.

\begin{figure}
\begin{center}
	\includegraphics[width=7cm]{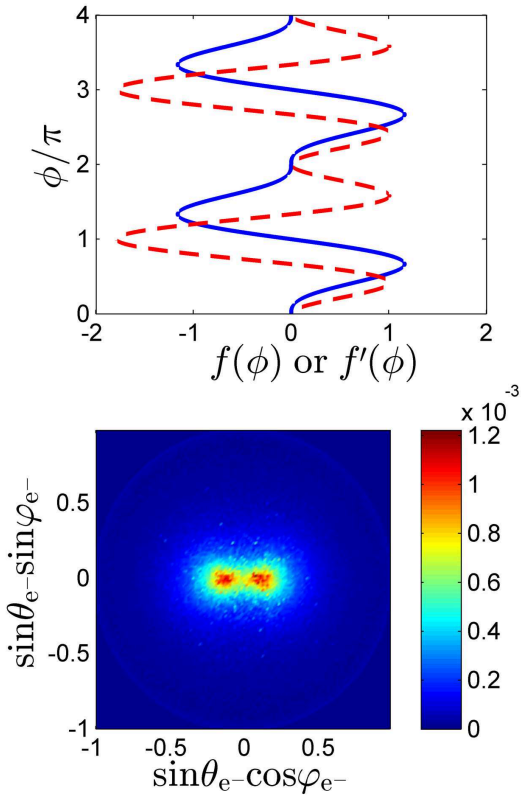}
	\end{center}
	\caption{(Color online) In the upper panel, we show shape functions for the four-vector potential 
$f(\phi)$ (blue solid line) and for the electric field $f'(\phi)$ (red dashed line), describing a bichromatic laser field defined 
by Eqs.~\eqref{t1} and~\eqref{t2}. The laser-field parameters are such that $\omega=m_{\rm e}c^2$ and $\mu=1$.
The lower panel presents coarse-grained angular distributions of probability rates of created electrons,
which have been projected onto the $xy$ plane.  
	\label{r1}}
\end{figure}
\begin{figure}
\begin{center}
	\includegraphics[width=7cm]{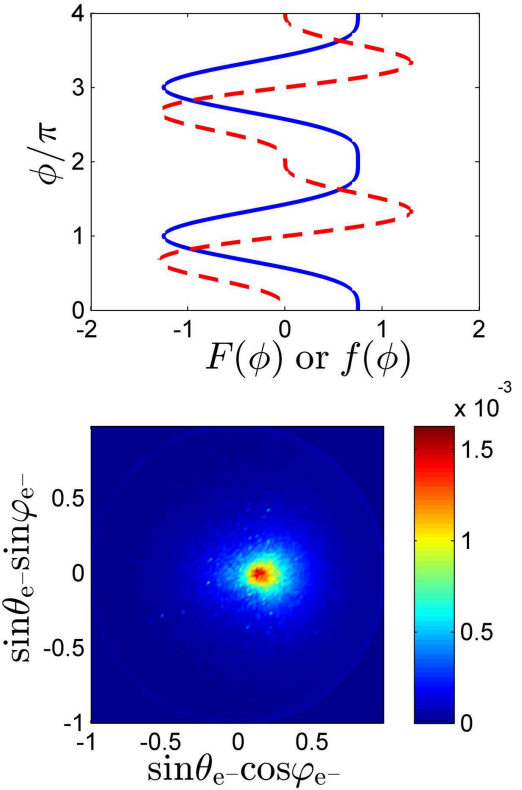}
	\end{center}
	\caption{(Color online) Same as in Fig.~\ref{r1} but  for the bichromatic laser field 
                 described by Eqs.~\eqref{t3} and~\eqref{t4}.
	\label{r2}}
\end{figure}

To be more specific, we consider here two distinct representations of the four-vector
potential which describe a two-color laser field. In the first case,
\begin{eqnarray}
 A(\phi)&=&A_0\varepsilon f(\phi)\,,\label{t1}\\
 f(\phi)&=&\frac{2}{\sqrt{5}}\Bigl[\sin(\phi)-\frac{1}{2}\sin(2\phi)\Bigr]\,,\label{t2}
\end{eqnarray}
whereas in the second case,
\begin{eqnarray}
 A(\phi)&=&A_0\varepsilon F(\phi)\,,\label{t3}\\
 F(\phi)&=&\frac{4}{\sqrt{17}}\Bigl[\cos(\phi)-\frac{1}{4}\cos(2\phi)\Bigr]\,,\label{t4}
\end{eqnarray}
with $\phi=k\cdot x$. Here, we have introduced $A_0$ which describes the strength of 
the laser field and is related to the relativistically invariant parameter $\mu=|eA_0|/(m_{\rm e}c)$. 
As anticipated above, the polarization and wave four-vectors are $\varepsilon=(0,{\bm e}_x)$ and $k=(\omega/c)(1, -{\bm e}_z)$, 
respectively. One can check that for this configuration, it holds that $k^2=0$ and $k\cdot A=0$. These
conditions constitute foundations of the theory developed in Ref.~\cite{b1}. Moreover, in Eqs.~\eqref{t2} and~\eqref{t4},
we have introduced the so-called shape functions $f(\phi)$ and $F(\phi)$, normalized such that 
\begin{equation}
 \langle f^2\rangle=\frac{1}{2\pi}\int_0^{2\pi}{\rm d}\phi f^2(\phi)=\frac{1}{2}\,,
\end{equation}
and similarly for $\langle F^2\rangle=1/2$. As we have discussed in~\cite{b1}, 
imposing these conditions guarantees that the ponderomotive energy of the electron
quiver oscillations in a laser field is the same in both cases. In other words, the
pair creation threshold energy is also the same, which makes it meaningful to compare
the rates of pair production for both considered cases. Let us note that for our choice of the
shape functions, it happens that $F'(\phi)\sim f(\phi)$. This means that while the 
shape function $f(\phi)$ describes the vector potential for the first choice [Eqs.~\eqref{t1} and~\eqref{t2}], 
it relates to the electric component of the laser field for our second choice [Eqs.~\eqref{t3} and~\eqref{t4}]. 
It is the purpose of this paper to investigate which of these factors actually determines the behavior 
of probability rates of the nonlinear Bethe-Heitler process.

In the lower panel of Fig.~\ref{r1}, we present coarse-grained angular maps of created electrons 
projected onto the $xy$ plane, for the case described by Eqs.~\eqref{t1} and~\eqref{t2} such that
$\omega=m_{\rm e}c^2$ and $\mu=1$. These results have been obtained using the Monte-Carlo method 
that we developed in~\cite{KK1}. Here, $3\times 10^9$ sample points were collected, which then 
were averaged according to the prescription described in Ref.~\cite{KK1}. We have also calculated
the total probability rate of pair creation which turned out to be 
$0.11\times 10^{-5}$ in relativistic units, with an estimated error less than $1\%$.
It is crucial to note that in the considered case, electrons (positrons as well) are created symmetrically 
with respect to the $x$ axis, i.e., with respect to the polarization vector ${\bm \varepsilon}$.
In the upper panel of Fig.~\ref{r1}, we demonstrate the shape functions $f(\phi)$ and $f'(\phi)$
which describe either the vector potential (blue solid line) or the electric field 
(red dashed line), respectively. There is a clear symmetry of $f(\phi)$ as opposed to an asymmetry of 
$f'(\phi)$. One can see therefore that the symmetry of the vector potential is reflected 
in the pattern of angular distributions of created particles.

Fig.~\ref{r2} relates to the second choice of the vector potential 
[see, Eqs.~\eqref{t3} and~\eqref{t4}], with the same laser-field parameters as in Fig.~\ref{r1}.
In this case, the total probability rate of pair creation is $0.46\times 10^{-6}$ in relativistic
units, with an error less than $1\%$. In the upper panel, we plot the shape functions $F(\phi)$ 
and $f(\phi)$. As we have already mentioned, in the current case $F(\phi)$ describes the vector potential 
(blue solid line) whereas $f(\phi)$ corresponds to the electric field (red dashed line).
This time, the vector potential shape function $F(\phi)$ is asymmetric while the
electric field shape function $f(\phi)$ stays symmetric. In the lower panel of Fig.~\ref{r2}, we
present the coarse-grained angular distributions of produced electrons. Clearly, an asymmetric 
pattern is observed there. Again, the behavior of the electric field component seems to
be irrelevant when it comes to properties of pair creation. The same will be 
elaborated below, when analyzing partial angular distributions of created particles.

\begin{figure}
\begin{center}
	\includegraphics[width=8.5cm]{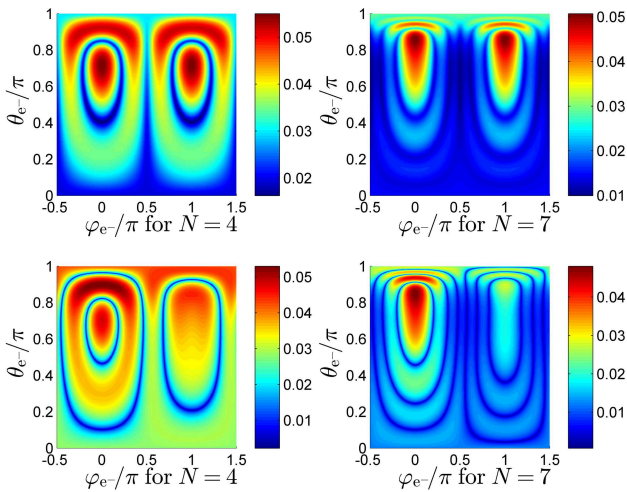}
	\end{center}
	\caption{(Color online) Angular maps of $R_4({\bm q}_{\rm f},\hat{\bm p}_{{\rm e}^-})$ (left column)
and $R_7({\bm q}_{\rm f},\hat{\bm p}_{{\rm e}^-})$ (right column) in the case when
${\bm q}_{\rm f}=-2m_{\rm e}c{\bm e}_z$. The parameters of the laser field have been chosen
such that $\omega=m_{\rm e}c^2$ and $\mu=1$. While the upper panels refer to the 
vector potential specified by Eqs.~\eqref{t1} and~\eqref{t2}, the lower panels are for the 
vector potential described by Eqs.~\eqref{t3} and~\eqref{t4}. For visual purposes, the rates 
have been raised to power 1/5.
	\label{r3}}
\end{figure}
\begin{figure}
\begin{center}
	\includegraphics[width=8.5cm]{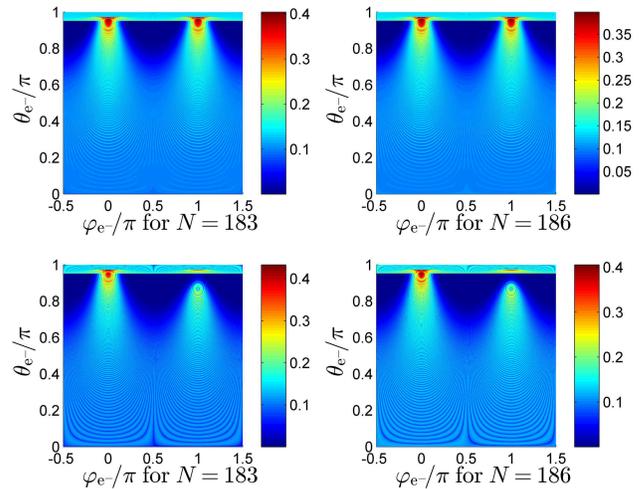}
	\end{center}
	\caption{(Color online) Angular maps of $R_{183}({\bm q}_{\rm f},\hat{\bm p}_{{\rm e}^-})$ (left column)
and $R_{186}({\bm q}_{\rm f},\hat{\bm p}_{{\rm e}^-})$ (right column) in the case when
${\bm q}_{\rm f}=-2m_{\rm e}c{\bm e}_z$.
The parameters of the laser field are $\omega=m_{\rm e}c^2$ and $\mu=10$. The upper panels 
correspond to the case specified by Eqs.~\eqref{t1} and~\eqref{t2}, whereas the lower panels are for the 
vector potential described by Eqs.~\eqref{t3} and~\eqref{t4}. For visual purposes, the rates 
have been raised to power 1/10.
	\label{r4}}
\end{figure}

In the following, $R_N^{(\ell)}({\bm q}_{\rm f},\hat{\bm p}_{{\rm e}^-})$ denotes the partial triply differential 
probability rate of pair creation when the value of energy transfer from the colliding 
proton as well as orientations of the proton and the emitted electron are fixed. In Refs.~\cite{b1,KK2,KK3}, we 
have also used the quantity $R_N({\bm q}_{\rm f},\hat{\bm p}_{{\rm e}^-})=
\sum_\ell R_N^{(\ell)}({\bm q}_{\rm f},\hat{\bm p}_{{\rm e}^-})$ which is a sum of the partial differential
probability rates for the same four-momenta conservation condition~\cite{KK2,KK3}. Thus in Figs.~\ref{r3} 
and~\ref{r4}, we present angular maps of $R_N({\bm q}_{\rm f},\hat{\bm p}_{{\rm e}^-})$
in function of the electron polar and azimuthal angles, $\theta_{{\rm e}^-}$ and $\varphi_{{\rm e}^-}$,
for chosen numbers of photons $N$ absorbed from the laser field (for more details, see Ref.~\cite{b1}).

In Fig.~\ref{r3}, we demonstrate angular maps of the triply differential probability rates of pair production 
by 4-laser photons $(N=4)$, $R_4({\bm q}_{\rm f},\hat{\bm p}_{{\rm e}^-})$, and by 7-laser photons
$(N=7)$, $R_7({\bm q}_{\rm f},\hat{\bm p}_{{\rm e}^-})$. The results are
for the case when the Doppler up-shifted frequency of the fundamental color field is $\omega=m_{\rm e}c^2$, 
$\mu=1$, and the proton recoil is ${\bm q}_{\rm f}=-2m_{\rm e}c{\bm e}_z$. While the upper panels of
Fig.~\ref{r3} are for the shape function $f(\phi)$ [Eq.~\eqref{t2}], the lower
panels are for the shape function $F(\phi)$ [Eq.~\eqref{t4}]. We observe that these angular maps
are stretched toward polar angles $\theta_{{\rm e}^-}$ close to $\pi$, which is in the propagation
direction of the laser field. This has been realized to be a general feature of strong-field
multiphoton processes in the relativistic domain, where the motion of electrons in the direction of the 
laser field propagation is driven by the magnetic component of the Lorentz force~\cite{b1,Klaiber,Mull}.
We note also that these angular
maps exhibit the same type of behavior as the vector potential does, namely, they are symmetric when the vector
potential is symmetric (upper panels), and asymmetric otherwise (lower panels).
In contrast to these very few photon processes, in Fig.~\ref{r4} we
present maps for $R_{183}({\bm q}_{\rm f},\hat{\bm p}_{{\rm e}^-})$ ($N=183)$ and 
$R_{186}({\bm q}_{\rm f},\hat{\bm p}_{{\rm e}^-})$ ($N=186$), as denoted below each panel. This time, 
the laser field is even stronger than in Fig.~\ref{r3}, since $\mu=10$. 
Even though in the current case, we basically reach the quasistationary
regime of pair creation, that is believed to occur when $\mu\gg 1$~\cite{Popov1,Rit,Mull,KK1}, 
we still observe the same symmetry properties of angular maps of created electrons as in Fig.~\ref{r3}.
These properties are governed by the behavior of the vector potential, not by the electric field.
If the Sauter-Schwinger mechanism~\cite{Sauter,Schwinger} was responsible for the nonlinear Bethe-Heitler process
this would not be the case. At this point, let us also mention that for this strong field (corresponding
to $\mu=10$) we reach the limit of computational capacity in our Monte Carlo simulations. For this reason,
we do not present the respective result for the total probability rate of pair creation.

Let us note that angular distributions of created particles through nonlinear Bethe-Heitler
scenario have been studied before for the case when an incident laser field is treated as a monochromatic plane 
wave (see, for instance, Refs.~\cite{KK2,KK4}). In this case, the calculated angular spectra turn out to be symmetric under 
a mirror reflection with respect to the polarization direction of the laser field.
At the same time, both the vector potential and the electric field are symmetric in the plane
wave approximation. By adding the second color field, one can break the symmetry of either
the vector potential or the electric field, and analyze how this may influence the pair production.
This is the essence of the present paper and also a suggestion for experimentalists to
validate our idea.

In summary, symmetry properties of nonlinear Bethe-Heitler process by a two-color
laser field have been investigated. These properties have been proved to depend on
the vector potential characterizing the laser field, not on the electric field component.
This can be understood if we note that in the strong-field QED formalism which is based on the so-called
Volkov waves~\cite{KK2,Volkov}, the maximum of probability rates of pair creation
is achieved for saddle-point solutions of the classical action describing a charged particle
in the laser field. In our case, these stationary points are approached when 
produced particles are detected with asymptotic momenta such that their component along the polarization direction is equal to $e{\bm A}$. 
Hence, the vector potential {\it does} determine properties of ultrastrong pair creation.
In this sense also, the Sauter-Schwinger mechanism which relies on the electric-field dependence
of pair creation probability rates appears to be irresponsible for the process
under investigations, at least for the chosen laser-field parameters.

This work is supported by the Polish National Science Center (NCN) under Grant No. 2011/01/B/ST2/00381.


\begin{thebibliography}{00}


\bibitem{Sauter}
F. Sauter, Z. Phys. {\bf 69}, 742 (1931).

\bibitem{Schwinger}
J. Schwinger, Phys. Rev. {\bf 82}, 664 (1951).

\bibitem{Ehlotzky}
C. Itzykson and J. -B. Zuber, {\it Quantum Field Theory} (McGraw-Hill, New York, 1980).

\bibitem{Brezin}
E. Brezin and C. Itzykson, Phys. Rev. D {\bf 2}, 1191 (1970).

\bibitem{Popov}
V. S. Popov, Zh. Eksp. Teor. Fiz. {\bf 61}, 1334 (1971) [Sov. Phys. JETP {\bf 34}, 709 (1972)].

\bibitem{Bula}
D. L. Burke, R. C. Field, G. Horton-Smith, J. E. Spencer, D. Walz, S. C. Berridge, W. M.
Bugg, K. Shmakov, A. W. Weidemann, C. Bula, K. T. McDonald, E. J. Prebys, C.
Bamber, S. J. Boege, T. Koffas, T. Kotseroglou, A. C. Melissinos, D. D. Meyerhofer,
D. A. Reis, and W. Ragg, Phys. Rev. Lett. {\bf 79}, 1626 (1997).

\bibitem{Bamber}  
C. Bamber, S. J. Boege, T. Koffas, T. Kotseroglou, A. C.
Melissinos, D. D. Meyerhofer, D. A. Reis, W. Ragg, C. Bula, K. T. McDonald,
E. J. Prebys, D. L. Burke, R. C. Field, G. Horton-Smith, J. E. Spencer, D. Walz,
S. C. Berridge, W. M. Bugg, K. Shmakov, and A. W. Weidemann, Phys. Rev. D {\bf 60}, 092004 (1999).

\bibitem{Mourou}
G. A. Mourou, T. Tajima, and S. V. Bulanov, Rev. Mod. Phys. {\bf 78}, 309 (2006).

\bibitem{b1}
K. Krajewska and J. Z. Kami\'nski, Phys. Rev. A {\bf 85}, 043404 (2012).

\bibitem{KK1}
K. Krajewska and J. Z. Kami\'nski, Phys. Rev. A {\bf 84}, 033416 (2011).

\bibitem{KK2}
K. Krajewska and J. Z. Kami\'nski, Phys. Rev. A {\bf 82}, 013420 (2010). 

\bibitem{KK3}
K. Krajewska, Laser Phys. {\bf 21}, 1275 (2011).

\bibitem{Klaiber}
M. Klaiber, K. Z. Hatsagortsyan, and C. H. Keitel, Phys. Rev. A {\bf 75}, 063413 (2007).

\bibitem{Mull}
C. M\"uller, K. Z. Hatsagortsyan, M. Ruf, S. J. M\"uller, H. G. Hetzheim, M. C. Kohler, and C. H. Keitel,
Laser Phys. {\bf 19}, 1743 (2009).

\bibitem{Popov1}
V. S. Popov, Zh. Eksp. Teor. Fiz. {\bf 63}, 1586 (1972) [Sov. Phys. JETP {\bf 36}, 840 (1973)].

\bibitem{Rit}
V. I. Ritus, J. Rus. Laser Res. {\bf 6}, 497 (1985).

\bibitem{KK4}
K. Krajewska and J. Z. Kami\'nski, Laser Phys. {\bf 18}, 185 (2008).

\bibitem{Volkov}
D. M. Volkov, Z. Phys. {\bf 94}, 250 (1935).


\end{thebibliography}
\end{document}